\documentclass[twocolumn,prl,aps,superscriptaddress,showpacs,preprintnumbers,amsmath,amssymb]{revtex4}
\usepackage{graphicx}

\begin{document}

\title{ Magnetic resonance in the model
high-temperature superconductor HgBa$_2$CuO$_{4+\delta}$ }

\author{G.~Yu}
\affiliation{Department of Physics, Stanford University, Stanford,
California 94305, USA}
\author{Y.~Li}
\affiliation{Department of Physics, Stanford University, Stanford,
California 94305, USA}
\author{E.M.~Motoyama}
\affiliation{Department of Physics, Stanford University, Stanford,
California 94305, USA}
\author{X.~Zhao}
\affiliation{Stanford Synchrotron Radiation Laboratory, Stanford,
California 94309, USA}
\affiliation{State Key Lab of Inorganic Syntesis and Preparative
Chemistry, College of Chemistry, Jilin University, 2699 Qianjin
Street, Changchun 130012, P.R. China}
\author{N.~Bari\v{s}i\'{c}}
\affiliation{Stanford Synchrotron Radiation Laboratory, Stanford,
California 94309, USA}
\affiliation{1. Physikalisches Institut, Universit\"{a}t Stuttgart,
Pfaffenwaldring 57, 70550 Stuttgart,Germany }
\author{Y.~Cho}
\affiliation{Stanford Synchrotron Radiation Laboratory, Stanford,
California 94309, USA}
\affiliation{Research Center for Dielectric and Advanced Matter
Physics, Pusan National University, Busan 609-735, Korea}
\author{P.~Bourges}
\affiliation{Laboratoire L\'{e}on Brillouin, CEA-CNRS, CEA-Saclay,
91191 Gif-sur-Yvette,
France}
\author{K.~Hradil}
\affiliation{
Institut f\"{u}r Physikalische Chemie, Universit\"{a}t G\"{o}ttingen,
37077 G\"{o}ttingen, Germany
}
\author{R.A.~Mole}
\affiliation{Forschungsneutronenquelle Heinz Maier-Leibnitz, 85747
Garching, Germany}
\author{M.~Greven}
\affiliation{Department of Applied Physics, Stanford University,
Stanford, California
94305, USA}

\date{\today}

\begin{abstract}
We present an inelastic neutron scattering study of the structurally simple single-layer compound 
HgBa$_2$CuO$_{4+\delta}$
close to optimal doping ($T_c \approx  96$ K).  
A well-defined antiferromagnetic resonance with energy $\omega_r = 56$ meV  ($\approx 6.8 \, k_BT_c$) is observed below the superconducting transition temperature $T_c$. 
The resonance mode is energy-resolution limited and exhibits an intrinsic momentum width of about 
$0.2\,\mathrm{\mathring{A}^{-1}}$, consistent with prior work on several other cuprates. 
However, the unusually large value of the mode energy implies a non-universal relationship between
$\omega_r$ and $T_c$ across different families of cuprates.
\end{abstract}

\maketitle

A strong piece of evidence relating magnetic excitations to the superconductivity 
in the high-$T_c$ cuprates comes from inelastic neutron scattering (INS) experiments that reveal
a well-defined antiferromagnetic (AF) excitation in  
YBa$_2$Cu$_3$O$_{6+\delta}$ (YBCO), 
Bi$_{2}$Sr$_{2}$CaCu$_{2}$O$_{8+\delta}$ (Bi2212), and Tl$_{2}$Ba$_{2}$CuO$_{6+\delta }$ (Tl2201) \cite{Rossat-MignodPhysica91, MookPRL93,FongPRL95, BourgesPRB96, FongNature99, HePRL01, HeScience02}. 
Near optimal doping, this `resonance' appears at $\omega_r =$ 5-6 $k_{\rm B}T_{\rm c}$, and its intensity follows a power-law like behavior below $T_c$.
The resonance has been associated with evidence of a bosonic coupling 
with the charge degrees of freedom: photoemission reveals a dispersion 
anomaly in the single-particle excitations \cite{Damascelli03} and optical conductivity 
indicates the existence of a bosonic mode in the optical self-energy below 
$T_{\rm c}$ \cite{HwangPRB07},
both observed near the energy of the resonance.

There exist two classes of theoretical models for the resonance.
Most theoretical approaches consider it a
collective spin excitation in the presense of strong electronic
correlations. These spin exciton models interpret the resonance as a $S=1$ bound
state below the threshold of the electron-hole excitation continuum 
\cite{EschrigAdvPhys06}. 
They give a good description of the unusual
`hour-glass' dispersion, centered at the energy $\omega_r$ and wavevector
${\bf Q}_{\rm AF}=(1/2,1/2)$, 
that was first established for YBCO \cite{HaydenNature04,PailhesPRL04,ReznikPRL04,StockPRB05}. 
A second type of theoretical approach,
motivated by an INS study that revealed an hour-glass dispersion in 
stripe-ordered  La$_{1.875}$Ba$_{0.125}$CuO$_{4+\delta }$ 
\cite{TranquadaNature04}, considers localized spin models that treat the resonance
as a magnon-like excitation in a magnetically-ordered phase \cite{Carlson2004,
Kruger2004,Seibold2006,Uhrig2004, Vojta2006}.
In these models, the dispersion results from the spin-waves emerging from spin stripes and from one-dimensional triplet excitations within spin-ladders. 

Detailed studies of the resonance have so far have been limited to the structurally complicated bilayer compounds YBCO and Bi2212, for which the large single crystals required for state-of-the-art INS experiments 
have been available. The bilayer structure, with two closely spaced CuO$_{2}$ layers per unit cell, splits the resonance into two modes with odd and even symmetry under the exchange of the two layers \cite{PailhesPRL03,PailhesPRL04,PailhesPRL06, CapognaPRB07}, complicating the interpretation of the data.
Results for single-layer compounds are potentially easier to interpret. 
Although no direct evidence for a resonance has been observed in 
La$_{2-x}$Sr$_{x}$CuO$_{4}$ (LSCO) \cite{ChristensenPRL04, VignolleNaturePhysics07}, 
the result for Tl2201 suggests that the resonance is a universal
property of the hole-doped systems, regardless of the number of
CuO$_{2}$ layers per unit cell \cite{HeScience02}. 
Unfortunately, for Tl2201 it has not been possible to reach the interesting underdoped regime and to prepare larger crystals, which has prevented a systematic INS study. 
It would therefore be highly desirable to investigate a single-layer system that  (1) is structurally simple, (2) exhibits a high $T_c$, (3) is relatively free of disorder effects, (4) allows access over a wide range of doping, and (5)  can be grown in form of sizable crystals.

In this Letter, we present INS results for the magnetic resonance of nearly optimally-doped 
HgBa$_2$CuO$_{4+\delta}$ (Hg1201).
Hg1201 exhibits the highest $T_c$ at optimal doping (onset $T_{\rm c} =$ 96-97 K) of all single-layer cuprates, possesses a simple tetragonal crystal structure with a large spacing between neighboring CuO$_{2}$ layers (Fig. 1(a)), and is thought to be free of substantial disorder effects  \cite{Bobroff1997, Eisaki2004}. Crystals with a mass of up to 1.2 g \cite{ZhaoAdvMat06} and tunable over  wide doping range ($p\approx$ 0.08-0.22 to date) \cite{Barisic2008} have recently become available, 
enabling state-of-the-art neutron scattering work \cite{Li08}. Consequently,  Hg1201 is the most promising cuprate for a quantitative experimental study of the magnetic resonance. We demonstrate the existence of the resonance, and hence provide further evidence that this magnetic mode is a universal feature and that
LSCO is not fully representative of the cuprates.
The resonance in Hg1201 occurs at $\omega_r = 56(1)$ meV, or $6.8(2) \, k_BT_c$, a value that is considerably larger than for YBCO, Bi2201 and Tl2212 near optimal doping. This observation 
demonstrates that, contrary to previous suggestions \cite{Dai2001, SidisPRL00, WilsonNature06, Hufner2008}, the resonance energy is not universally related to $T_c$ across different families of materials.

The experiment was conducted on the thermal
triple-axis spectrometer PUMA at the FRM-II reactor in Garching, Germany. The sample consisted of $24$ co-mounted crystals with a total mass of about $1\,\mathrm{g}$. The growth involved a 
two-step flux method which yields underdoped samples \cite{ZhaoAdvMat06}. 
To achieve optimal doping, the crystals were subsequently annealed 
for about two months 
in air at $350^{\circ }\mathrm{C}$ to increase the oxygen concentration in the Hg layers, 
and hence the hole concentration in the Cu-O layers. The $T_{\rm c}$ of individual crystals was determined from DC susceptibility measurements [Fig. 1(c)], 
and the entire sample had an average onset transition of 
about 96 K. 
Prior to the experiment, the 
crystals were co-mounted on three aluminum plates and aligned using Laue X-ray diffraction [Fig. 1(b)]. 
The aluminum plates were subsequently co-aligned in the neutron beam such that the 
 (110) and (001) 
nuclear reflections were within the horizontal scattering plane (room temperature lattice
constants: $a,b=3.85\,\mathrm{\mathring{A}}$ and $c=9.5\,\mathrm{\mathring{A}}$).  
A mosaic scan at 
the (004) reflection indicated that the [001] crystallographic axes were aligned within 
$1.4^{\circ }$ [Fig. 1(d)]. 
The experiment involved pyrolytic graphite (002) reflections
for the double-focusing monochromator and analyzer, a fixed final neutron energy of 
30.5 meV, and no horizontal collimations. This configuration resulted in a momentum resolution of $0.1$ r.l.u. and an energy resolution of $8 \,\mathrm{meV}$ at 
energy transfers in the 50-60 meV range. 

\begin{figure}[tb]
\includegraphics[width=3.4in]{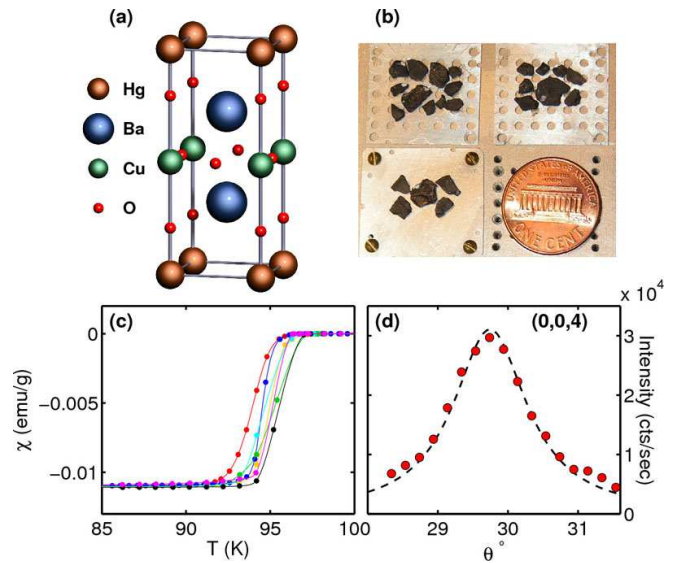}
\caption{(a) Schematic tetragonal (space group P4/mmm) crystal structure  of HgBa$_2$CuO$_4$ (without interstitial oxygen atom in the Hg layer). (b) Co-aligned crystals. 
(c) Magnetic susceptibility for seven representative crystals. (d) Rocking scan
through $(004)$ Bragg peak.}
\label{fig1}
\end{figure}

\begin{figure}[tb]
\includegraphics[width=2.8in]{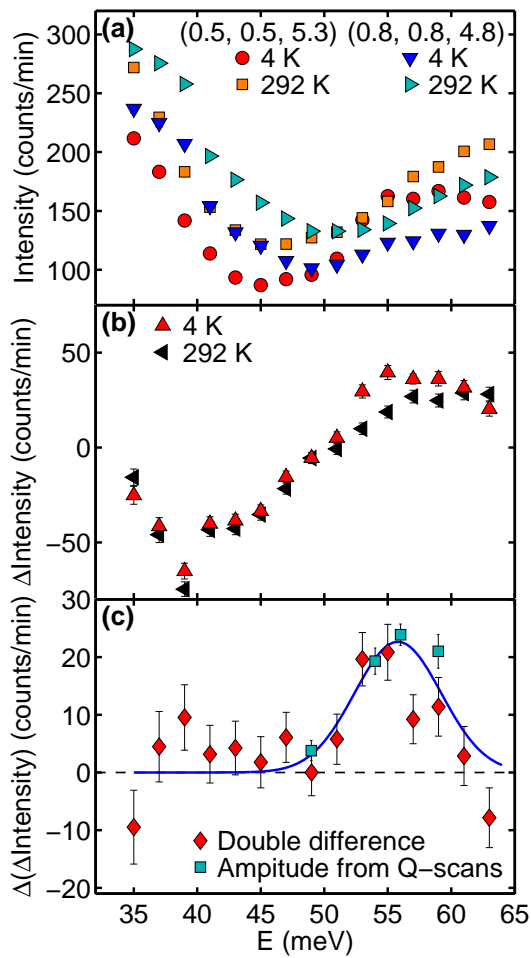}
\caption{(a) Energy scans at ${\bf Q}_{AF} = (0.5, 0.5, 5.3)$ and ${\bf Q}_{BG} =
(0.8, 0.8, 4.8)$, measured at 4 K and 292 K. (b)
The difference between the intensities at ${\bf Q}_{AF}$ and ${\bf Q}_{BG}$ shown in (a) 
reveals a relative enhancement of scattering at 4 K. (c) The double
difference reveals a well-defined (resolution-limited) enhancement of scattering at ${\bf Q}_{AF}$ upon cooling into the SC state. Phonon scattering, unlike magnetic scattering due to the resonance, is expected to decrease, not increase upon cooling. 
The square symbols are amplitude values at the two-dimensional zone center obtained from rocking scans at 4 K [see Fig. 3(a)]. The continuous line is the result of a fit to a resolution-limited Gaussian (FWHM of 8 meV) centered at  $\omega = 56(1)$ meV.
}
\label{fig2}
\end{figure}

Figure~\ref{fig2}(a) shows energy scans at 4 K and 292 K at the two-dimensional AF Brillouin zone center ${\bf Q}_{AF}=(0.5,0.5,5.3)$ and at a background position, ${\bf Q}_{BG}=(0.8,0.8,4.8)$. The latter was chosen so that  $|{\bf Q}_{BG}| = |{\bf Q}_{AF}|$ to minimize changes in phonon contributions, which have a relatively strong $Q$-dependence. The main background contributions are then largely removed by taking the difference of intensities at these two wavevectors.
Figure 2(b) reveals that the background-subtracted response near 56 meV is enhanced at low temperature. The underlying change upon cooling into the SC state is more clearly evident from Fig. 2(c), which shows the double difference of the scattering intensity and reveals a   
resolution-limited peak at 56(1) meV. This intensity enhancement at low temperature and ${\bf Q} = {\bf Q}_{AF}$ is the magnetic resonance. 

Further evidence for a well-defined excitation at ${\bf Q} = {\bf Q}_{AF}$ and $\omega_r = 56$ meV comes from constant-$Q$ scans such as those at 48 and 56 meV shown in Fig. 3(a). 
These `rocking' scans are scans in ${\bf Q}$-space 
through the two-dimensional AF zone center at fixed momentum transfer $|{\bf Q}| = |{\bf Q}_{AF}|$, 
which minimizes the slope of the background scattering.
As can be seen from Fig. 2(c), the amplitude obtained from Gaussian fits to such scans agrees well with the double-difference obtained from energy scans. 
The latter effectively removes phonon contributions as well as any additional weakly temperature-dependent magnetic contributions. 
We note that the thermal (Bose) population factor for $\omega = 56$ meV is only 1.12 at $T=292$ K and can thus be neglected in the above analysis.

For  Bi2212, the energy (and momentum) width of the resonance is rather large, and it has been suggested that this may be the result of the relatively large amount of quenched disorder present in this compound \cite{FongNature99, HePRL01,Fauque07}. 
On the other hand, 
as for optimally-doped YBCO \cite{Woo06} and Tl2201 \cite{HeScience02}, we find that the resonance in Hg1201 is energy-resolution-limited [see Fig. 2(c)], consistent with expectations that disorder effects are weak \cite{Bobroff1997,Eisaki2004}.


As shown in Fig. 3(b), the response at $\omega_r = 56$ meV remains commensurate at higher temperatures. As for optimally-doped YBCO, the measured intensity increases quickly below $T_c$. 
Figure 4 summarizes the temperature dependence of the scattering at $\omega = \omega_r$ and ${\bf Q} = {\bf Q}_{AF}$, as obtained from fits of momentum scans to a Gaussian (see Fig. 3).
A heuristic fit of the peak susceptibility to a power-law (`order parameter') form with Gaussian distribution of characteristic temperatures results in an estimated onset of the 
enhancement at 95.5(3.0) K, consistent with the average $T_c$ value of our sample. 

\begin{figure}[tb]
\includegraphics[width=2.8in]{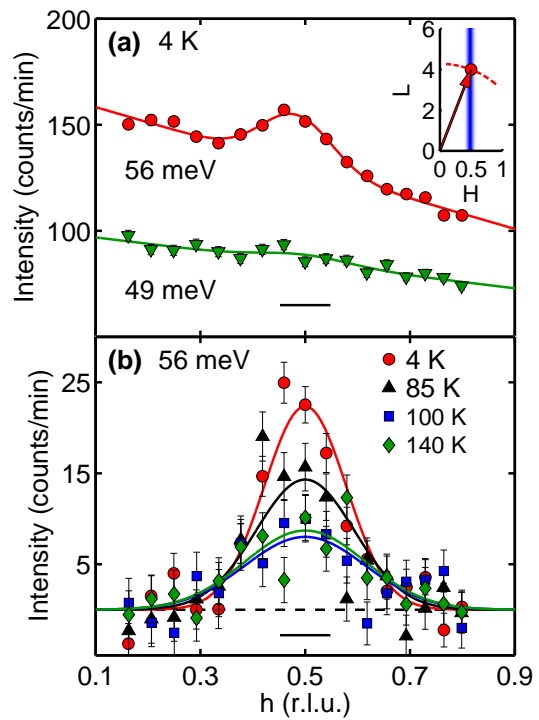}
\caption{(a) Constant-$Q$ (`rocking') scans across $(0.5, 0.5, 4)$ at energy transfers of 49 and 56 meV and a temperature of 4 K. The scan trajectory is indicated in the inset. (b) Rocking scans at four temperatures at the resonance energy $\omega_{\rm r} = 56$ meV, after subtraction of a linear background. The lines in (a) and (b) are the results of fits to a Gaussian. The horizontal bars indicate the FWHM momentum resolution.
}
\label{fig3}
\end{figure}

The response does not vanish completely above $T_c$. About one third of 
of the low-temperature intensity remains in the normal state, without any further temperature dependence up to 190 K. While this is different from prior results for YBCO \cite{BourgesPRB96}, a non-zero intensity centered at ${\bf Q} = {\bf Q}_{AF}$ is not unexpected. Even in the absence of the resonance for temperatures above $T_c$, the non-zero energy and momentum resolution of the experiment results in an overlap with the expected normal state magnetic excitations. Furthermore, part of the remaining signal could be due to phonons. 

The response at $\omega = \omega_r$ is broader than the momentum resolution, with an intrinsic width of approximately $0.2\,\mathrm{\mathring{A}^{-1}}$ (FWHM) deep in the SC state at 4 K. This value is consistent with the prior results for optimally-doped YBCO \cite{Rossat-MignodPhysica91, MookPRL93,
FongPRL95, BourgesPRB96} and Tl2201 \cite{HeScience02}. 
Figure 4(b) shows that the width obtained from Gaussian fits of the rocking scans 
does not exhibit an anomaly across $T_c$, consistent with a lack of a significant increase of
damping in the normal state, in contrast to predictions based on localized spin models \cite{Andersen2005}.
Nevertheless, our result is consistent with a monotonic increase of the momentum width with increasing temperature. The scattering near ${\bf Q} = {\bf Q}_{AF}$ may contain two distinct contributions in the SC state: (1) a broad and (nearly) temperature independent part already present in the normal state, and (2) the resonance, which is only present in the SC state.


Contrary to prior suggestions \cite{Dai2001, SidisPRL00, WilsonNature06, Hufner2008}, we conclude that the resonance energy does not universally depend on $T_c$. 
In overdoped YBCO  ($T_c = 75$ K) \cite{PailhesPRL06} and Bi2212  ($T_c = 70$ K) \cite{CapognaPRB07} the two mode energies were found to be equal within error: $\omega_r =$ 34-35 meV or about  $5.5\,k_B T_c$. Closer to optimal doping, the two mode energies begin to differ from each other in these double-layer compounds. Considering the average mode energy, one finds $\omega_r = $5.4-6.3 $\,k_B T_c$ near optimal doping \cite{PailhesPRL03,PailhesPRL04,PailhesPRL06,CapognaPRB07}.
For optimally-doped Tl2201 ($T_c =  92.5$ K), $\omega_{\rm r} = 47$ meV or $5.9\,k_B T_c$. 
While these values are overall consistent with each other, 
they are considerably smaller than our result of
$\omega_r = 6.8(2)\,k_B T_c$ for (nearly) optimally-doped Hg1201. 

We note that although there exists no universal relationship between $\omega_r$ and $T_c$ across different families of compounds, the odd-parity mode energy of YBCO appears to follow such a linear relationship as a function of doping \cite{PailhesPRL06}. Whether the doping dependence of the resonance mode energy in Hg1201 is simply determined by the value of $T_c$ remains an open question.

\begin{figure}[tb]
\includegraphics[width=3.0in]{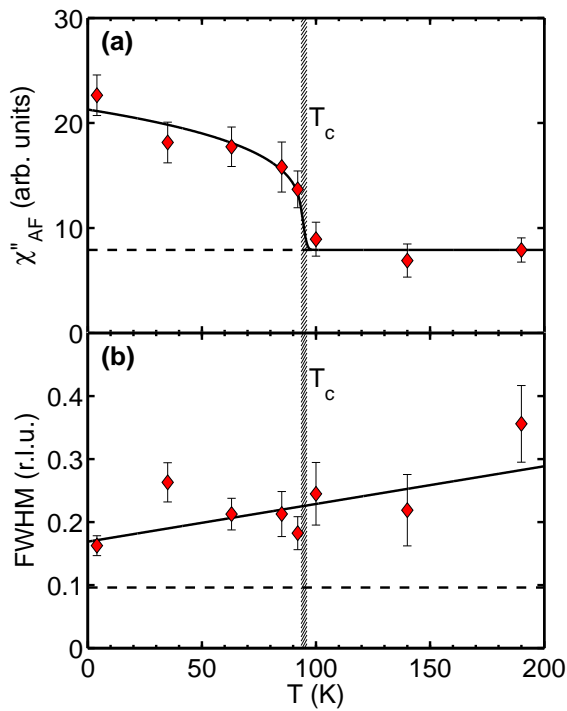}
\caption{Temperature-dependence of (a) peak susceptibility
(b) extrinsic momentum width obtained from Gaussian fits of rocking scans
[Figure \protect\ref{fig3}(b)] at $E_r=56$ meV. 
A fit to the heuristic form $\protect\chi _{AF}^{\prime \prime }\sim
(T_r-T)^{2\protect\beta }$, with small Gaussian distribution of
the characteristic temperature $T_r$, gave $T_{\mathrm{r}%
}=95.5(3.0)\,\mathrm{K}$, consistent with $T_{\mathrm{c}}=96(1)\,\mathrm{K}$%
, and $\protect\beta =0.12(3)$. The low-temperature ($4\,\mathrm{K}$)
extrinsic width of $0.16(2)\,\mathrm{r.l.u.}$ corresponds to an intrinsic
width of about $0.2\,\mathrm{\mathring{A}^{-1}}$. The solid line in (b) is the
result of a linear fit, whereas the dashed horizontal line indicates the instrument resolution.}
\label{fig4}
\end{figure}

In summary, we have performed a neutron scattering study of the magnetic resonance in Hg1201 close to optimal doping. Overall, the signatures of the resonance in this system are remarkably similar to double-layer YBCO and single-layer Tl2201. 
However, our results imply that the resonance energy does not scale universally with the superconducting transition temperature. 
The present work lays the foundation for the systematic study of the doping dependence of the magnetic resonance 
and, more generally, of the full dynamic magnetic susceptibility in this model single-layer copper-oxide superconductor.

This work was supported by the DOE under Contract No. DE-AC02-76SF00515 and by the NSF under Grant. No. DMR-0705086.

\bibliography{Hg1201_PRL_condmat}

\end{document}